\documentclass[twocolumn]{aastex631}

\hypersetup{linkcolor=red,citecolor=blue,filecolor=cyan,urlcolor=blue}

\usepackage{graphicx,color}
\usepackage{amssymb}
\usepackage{amsmath}
\usepackage{url}
\usepackage{natbib}
\usepackage{txfonts}
\usepackage{multirow}
\usepackage{array}
\usepackage{rotating}
\usepackage{sidecap}
\usepackage{hyperref}
\usepackage{epstopdf}
\usepackage{footnote}
\usepackage{tabularx}
\usepackage{booktabs}
\usepackage{longtable}
\usepackage{tabu}
\usepackage{longtable}

\newcommand{\SiIV}{\ion{Si}{4}}

\newcommand{\MgII}{\ion{Mg}{2}}

\newcommand{\Halpha}{H$\alpha$}

\newcommand{\kms}{km~s$^{-1}$}

\newcommand{\hinode}{\textit{Hinode}}

\newcommand{\sdo}{\textit{SDO}}
\newcommand{\iris}{\textit{IRIS}}

\begin{document}
		\title{ Genesis and  Coronal-jet-generating Eruption of a Solar Minifilament Captured by IRIS Slit-raster Spectra}

\correspondingauthor{Navdeep K. Panesar}
\email{panesar@lmsal.com}

\author[0000-0001-7620-362X]{Navdeep K. Panesar}
\affil{Lockheed Martin Solar and Astrophysics Laboratory, 3251 Hanover Street, Bldg. 252, Palo Alto, CA 94304, USA}
\affil{Bay Area Environmental Research Institute, NASA Research Park, Moffett Field, CA 94035, USA}

	\author[0000-0001-7817-2978]{Sanjiv K. Tiwari}
\affil{Lockheed Martin Solar and Astrophysics Laboratory, 3251 Hanover Street, Bldg. 252, Palo Alto, CA 94304, USA}
\affil{Bay Area Environmental Research Institute, NASA Research Park, Moffett Field, CA 94035, USA}

\author[0000-0002-5691-6152]{Ronald L. Moore}
\affil{Center for Space Plasma and Aeronomic Research (CSPAR), UAH, Huntsville, AL 35805, USA}
\affil{NASA Marshall Space Flight Center, Huntsville, AL 35812, USA}

\author[0000-0003-1281-897X]{Alphonse C. Sterling}
\affil{NASA Marshall Space Flight Center, Huntsville, AL 35812, USA}

\author[0000-0002-8370-952X]{Bart De Pontieu}
\affil{Lockheed Martin Solar and Astrophysics Laboratory, 3251 Hanover Street, Bldg. 252, Palo Alto, CA 94304, USA}
\affil{Rosseland Centre for Solar Physics, University of Oslo, P.O. Box 1029 Blindern, NO–0315 Oslo, Norway}
\affil{Institute of Theoretical Astrophysics, University of Oslo, P.O. Box 1029 Blindern, NO–0315 Oslo, Norway}

\begin{abstract}
We present the first \iris\  \MgII \ slit-raster spectra that fully capture the genesis and coronal-jet-generating eruption of a central-disk solar minifilament.  The minifilament arose in a negative-magnetic-polarity coronal hole.  The \MgII\ spectroheliograms verify that the minifilament plasma temperature is chromospheric.  The \MgII\ spectra show that the erupting minifilament’s plasma has blueshifted upflow in the jet spire’s onset and simultaneous redshifted downflow at the location of the compact jet bright point (JBP).   From the \MgII\ spectra together with AIA EUV images and HMI magnetograms, we find: (i) the minifilament forms above a flux-cancelation neutral line at an edge of a negative-polarity network flux clump; (ii) during the minifilament’s fast-eruption onset and jet-spire onset, the JBP begins brightening over the flux-cancelation neutral line.  From IRIS${^2}$ inversion of the \MgII\ spectra, the JBP’s \MgII\  bright plasma has electron density, temperature, and downward (red-shift) Doppler speed of 10$^{12}$  cm$^{-3}$, 6000 K, and 10 \kms, respectively, and the growing spire shows clockwise spin.  We speculate: (i) during the slow rise of the erupting minifilament-carrying twisted flux rope, the top of the erupting flux-rope loop, by writhing, makes its field direction opposite that of encountered ambient far-reaching field; (ii) the erupting kink then can reconnect with the far-reaching field to make the spire and reconnect internally to make the JBP.  We conclude that this coronal jet is normal in that magnetic flux cancelation builds a minifilament-carrying twisted flux rope and triggers the JBP-generating and jet-spire-generating eruption of the flux rope.
	\end{abstract}

\keywords{Solar magnetic fields (1503); Solar coronal holes (1484); Solar ultraviolet emission (1533); Solar magnetic reconnection (1504); Spectroscopy (1558); Jets (870); Solar corona (1483); Solar chromosphere (1479)}

\section{Introduction} \label{sec:intro}

Solar jets of all sizes are transient eruptive events. They appear as narrow structures that extend outward from the solar surface into the corona \citep{shibata11,raouafi16,innes16,shen21,schmieder22}.  They have been often observed in ultraviolet (UV;   \citealt{pike98,lei19,zhang21,joshi21,schmieder21}), extreme ultraviolet (EUV; \citealt{ywang98,nistico09,schmieder13,panesar16a,sterling17}), and  X-ray images \citep{shibata92,yokoyama95,alexander99,moore18,kyoung-sun20}. Coronal jets typically show a compact base-edge brightening and a bright spire during the eruption onset \citep{shibata92}. The base-edge brightening is called the jet  bright point (JBP; \citealt{sterling15}). Jets occur all over the solar disk and at the solar limb. They are numerous in the polar regions ($\sim$ 60 day$^{-1}$,  from \hinode\ soft X-ray images; \citealt{savcheva07,cirtain07}).
From a study of polar coronal hole jets in X-rays, \cite{savcheva07} found that their typical lifetime is $\sim$ 10 minutes, typical width is $\sim$ 8000 km and typical length is $\sim$ 5$\times$ 10$^{4}$ km.

With the availability of high spatial resolution extreme ultraviolet (EUV) images from  Solar Dynamics Observatory (SDO; \citealt{pensell12})/Atmospheric Imaging Assembly (AIA; \citealt{lem12}), many single coronal jets studies \citep[e.g.][]{hong11,shen12,adams14,solanki19,mazumder19_MFjetcancel} and several multiple jet studies \citep[e.g.][]{sterling15,sterling17,panesar16b,panesar18a,mcglasson19} reported that jets are  driven by a minifilament eruption. The JBP often appears near or at the site of the erupting minifilament \citep{sterling15}.  The length of pre-jet minifilaments is in the range of 10 -- 20 $\times$ 10$^{3}$ km \citep{wang00,sterling15,panesar16b}, which is significantly smaller than the size of filaments (about 3$\times$10$^{4}$ km -- 1.1$\times$10$^{5}$ km; \citealt{bernasconi05}). The pre-jet minifilament lies  in highly sheared magnetic field along a magnetic neutral line (also known as a  polarity inversion line; \citealt{martin86}) in a solar filament channel as do all typically-sized solar filaments \citep{martres66,gai97,martin98,mackay10}.

Several single-jet studies have found  evidence of flux cancelation at the base of coronal jets before and during the jet \citep[e.g.][]{huang12,young14a,young14b,hong19}. Recently, \cite{panesar16b,panesar17,panesar18a} investigated the formation and eruption of the minifilament in 10 on-disk quiet-region and 13 on-disk coronal-hole jets.  
They found that each  pre-jet minifilament formed by magnetic flux cancelation at the minifilament's neutral line. Further flux cancelation  at the neutral line triggers the minifilament eruption that drives the coronal jet. Later, \cite{mcglasson19} studied 60 coronal jets and found that at least 85\% of them result from minifilament eruption triggered by magnetic flux cancelation. Consistent results of flux convergence and cancelation in many coronal hole jets were reported by \cite{muglach21}. Similar  flux cancelation before and during eruption has also been observed for some active region jets \citep{sterling16,sterling17,mulay16,yang19_MFjetcancel,yang20_ARjetCan,poisson20_ARjetCan,zhangYJ21_ARjetCan}.  Other ideas for driving jets include flux emergence \citep{shibata92,yokoyama95}, alfv\'{e}nic
	magnetic twist-wave process without a flux rope \citep{pariat09} and building the explosive sheared field by photospheric shearing flows \citep{kumar18,kumar19}.

Small-scale UV/EUV jets (jetlets; \citealt{raouafi14})  occur at the edges of magnetic network lanes \citep{panesar18b,panesar19,panesar20b}. Although jetlets were initially found to concentrate at the base of plumes \citep{raouafi14}, they were later observed to occur at several locations on the Sun and not limited to the plume regions \citep{panesar18b}. Further \cite{panesar18b} argued that jetlets are miniature versions of coronal jets because: (i) they occur at neutral lines at the edges of magnetic  network lanes; (ii) they show base brightening on the neutral line during the eruption onset, and (iii) flux cancelation at the neutral line leads to the  eruption. However, in the UV and EUV images of the UV/EUV jetlets studied by \cite{panesar18b,panesar19,panesar20b}, no cool-plasma erupting minifilament-like dark feature is unambiguously seen in any jetlet's base before and during the jetlet's eruption onset. Nonetheless, \cite{panesar18b} observed twisting motion in one of the jetlets spire, which could be the result of a erupting flux rope from the jetlet base.  
On average, jetlets are three times smaller ($\sim$ 5000 km; \citealt{panesar18b}) in the base widths than typical coronal jets ($\sim$ 18000 km; \citealt{panesar16b}). Their duration is three to four times shorter (3 minutes) than the typical duration of coronal jets ($\sim$ 10 minutes; \citealt{savcheva07}).

Small-scale filaments have been previously observed and reported using \Halpha\ data, and were proposed to be miniature versions of large-scale solar filaments \citep{hermans86,wang00,lees03}. However, these studies did not have spectroscopic observations to infer temperatures, densities, and Doppler velocities of minifilaments. Further, these investigations did not explore whether typical coronal jets were driven by the eruptions of their small-scale filaments.

Here, we report  observations of a network-edge  coronal hole jet, made by the eruption of a  minifilament, observed  by \iris\  \MgII \ spectra. The observations are UV data from the \textit{Interface Region Imaging Spectrograph} (\iris; \citealt{pontieu14}), EUV images from \sdo/AIA, and line of sight magnetograms from \sdo/Helioseismic
and Magnetic Imager (HMI; \citealt{scherrer12}). To the best of our knowledge, this is the most detailed analysis so far of an on-disk minifilament eruption and coronal jet that is so fully captured by \iris\  \MgII \ spectra raster scans and spectroheliograms and is conspicuous in both. Our purpose is to investigate the properties of a coronal jet (including the pre-jet minifilament formation and evolution, the minifilament eruption, and the JBP) using \iris\  \MgII\ k spectra, and compare these with those observed in simultaneous EUV observations of \sdo/AIA.


%
\begin{figure*}
	\centering
	\includegraphics[width=\linewidth]{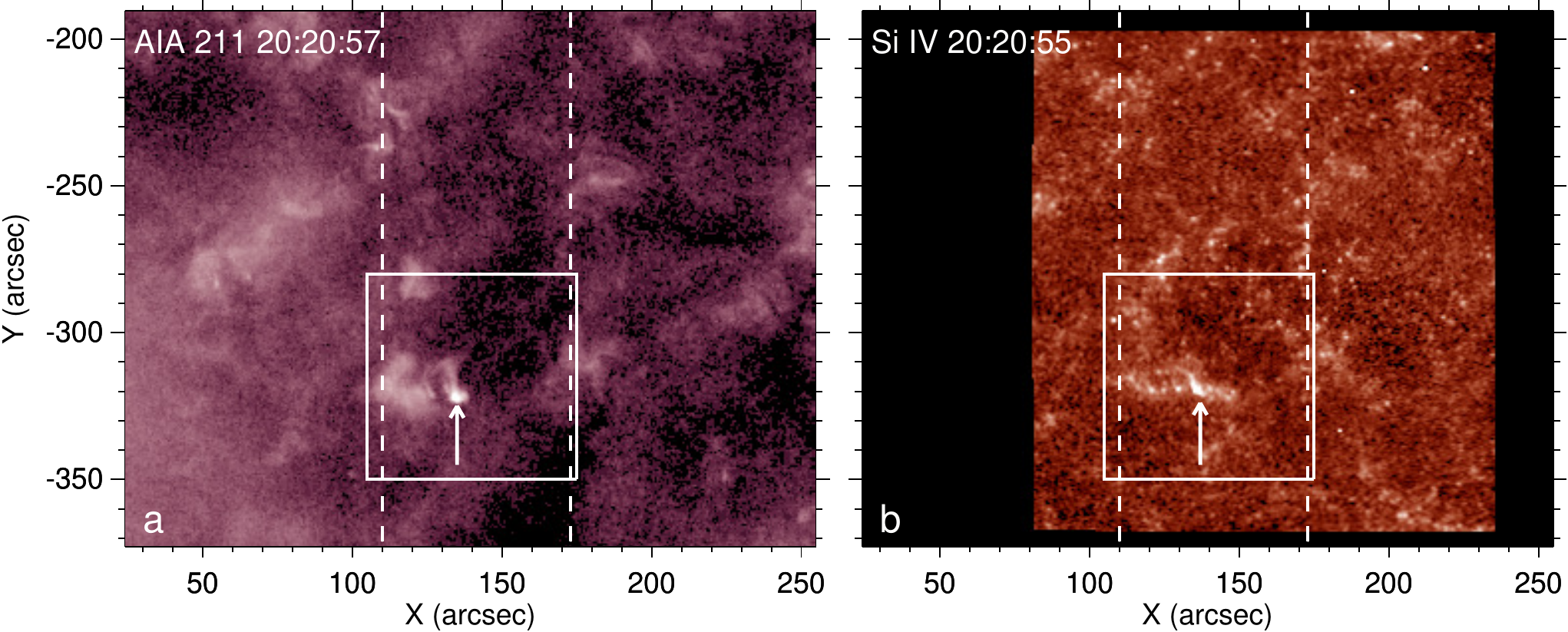} 
	\caption{Coronal jet observed in a  coronal hole on 08-April-2020. Panels (a) and (b), respectively, are an AIA 211 \AA\ image and an \iris\  1400 \AA\ SJI of the coronal hole during the jet. The white arrow points to the jet base brightening site. The white box outlines the FOV in Figure \ref{fig2}. The two vertical dashed lines bound the area that is scanned by the \iris\ slit.
	}  \label{fig1}
\end{figure*} 
%
%
\section{Observations}\label{data} 
We searched through the \iris\ data library for a jet that was well scanned by the \iris\ slit. After checking a large sample of jets  captured by \iris, we found a jet that is in an on-disk coronal hole and  is nicely scanned by the \iris\ slit. We verified the presence of the jet in \sdo/AIA 171 \AA\ data using JHelioviewer software \citep{muller17}. In most other cases, either the IRIS slit missed the jet or only a part of the jet was scanned. The  jet presented here is exceptional in that its evolution was fully captured under the \iris\ slit.

For our analysis, we mainly used data from \iris\ \citep{pontieu14}. \iris\  observed in its slit-jaw \MgII\ and \SiIV\ filters, an on-disk coronal hole region in a field of view (FOV) of 231\arcsec\ $\times$ 175\arcsec, on 08-April-2020  during 17:14 -- 22:17 UT. It ran a very large sparse 64-step raster with a step cadence of 1.1 s and a raster cadence of 72 s, making a total of 250 rasters (OBS ID 3645202860).  The width of the \iris\ slit is 0.33\arcsec. The data is summed along the slit by a factor of four, therefore making the pixels of size 0.33\arcsec $\times$ 0.66\arcsec, instead of 0.33\arcsec\ $\times$ 0.166\arcsec\ (unsummed).  The spectrograph slit step size is 1\arcsec.  \iris\ also took slit-jaw images (SJIs) in three different channels (\SiIV\ 1400 \AA, \MgII\ k 2796 \AA, and \MgII\ wing 2832 \AA) with a cadence of 3 seconds. For our investigation, we mainly focus on the \MgII\ k line, which samples chromospheric plasma, to study the detailed dynamics of the filament and jet. 
Because the signal to noise ratio of \SiIV\ 1400 \AA\ SJIs, which see transition region plasma, was low (and barely show some of the jet structure), we only used it as a context image in Figure \ref{fig1}. We summed two consecutive \SiIV\ 1400 \AA\ SJIs to enhance the signal. The \iris\ observing sequence was specifically chosen to capture the chromospheric aspects of the eruption at a high cadence, necessitating very short exposure times, and thereby sacrificing the  signal to noise in the far-ultraviolet (both in SJIs and spectra).	
We used level 2 \iris\ data\footnote{\url{https://www.lmsal.com/hek/hcr?cmd=view-event\&event-id=ivo\%3A\%2F\%2Fsot.lmsal.com\%2FVOEvent\%23VOEvent_IRIS_20200408_171435_3645202860_2020-04-08T17\%3A14\%3A352020-04-08T17\%3A14\%3A35.xml}} that were already  calibrated: dark current subtracted, flat-fielded, and corrected for geometrical distortion. 
  
%

We also used 304, 171, 193, and 211 \AA\ EUV images from \sdo/AIA  to study the filament  and jet in transition region and coronal emission \citep{lem12}. The  jet is definitely a  coronal jet because it is visible in the hotter  AIA channels, e.g. 335 \AA\ and 94 \AA.  \sdo/AIA provides full-disk solar images in seven different EUV channels  having a pixel size of 0.6\arcsec\ and a temporal cadence of 12 s. To study the photospheric roots of the magnetic field in and around  the jet-base region, we used line of sight magnetograms from \sdo/HMI \citep{scherrer12}. The HMI magnetograms have 0.5\arcsec\ pixels, 45 s temporal cadence and a noise level of about 7 G \citep{schou12,couvidat16}. AIA and HMI datasets were downloaded from the JSOC website\footnote{http://jsoc.stanford.edu/ajax/exportdata.html}. \iris\ and \sdo\ data sets were co-aligned and analyzed using SolarSoft routines \citep{freeland98}. Here, we show images from the  AIA 171 and 193 \AA\ channels because these two channels best show the minifilament and jet.

\floattable
	\begin{center}
		\begin{table*}
		\caption{Minifilament and Jet Times and Properties \label{tab:list}}
		\renewcommand{\arraystretch}{1.0}
		\begin{tabular}{c*{2}{c}}
			\noalign{\smallskip}\tableline\tableline \noalign{\smallskip}
		
			
			  Minifilament slow-rise start time in AIA 171\AA\ &  20:10 UT   \\ 
		Start time of western lobe brightening &  20:11 UT  \\
			  Jet bright point (JBP) start time in both AIA 171\AA\ and  IRIS \MgII\ k line &  20:18  UT  \\
			  Start time of eastern remote brightening &  20:20 UT  \\
	        Jet spire start time in both AIA 171\AA\ and  IRIS \MgII\ k line  &  20:20  UT    \\
		Jet spire duration in both AIA 171\AA\ and  IRIS \MgII\ k line &  10  min     \\   
	 Minifilament's slow-rise plane-of-sky speed & 3.0 $\pm$ 0.8 \kms	\\
			 Jet spire's plane-of-sky speed in AIA 171 \AA\   &  30 $\pm$ 4 \kms    \\  
		 Jet spire's plane-of-sky speed in IRIS \MgII\ k line &  32 $\pm$ 4 \kms   \\  
		Positive-polarity flux cancelation rate & 0.33 $\times$ 10$^{18}$ Mx hr$^{-1}$  \\   
		Negative-polarity flux reduction rate &  0.9 $\times$ 10$^{18}$ Mx hr$^{-1}$  \\  
		\noalign{\smallskip}\hline \noalign{\smallskip} 
	Physical properties obtained from the	IRIS$^{2}$ inversion of \MgII\ spectra:  \\
			\noalign{\smallskip}\hline \noalign{\smallskip}
			Electron density of JBP & 10$^{12}$ cm$^{-3}$ \\ 
			Temperature of JBP & 6000 K \\
			Doppler speed of JBP & 10 \kms\ (redshift) \\
			Upward Doppler speed in the jet spire's southern edge & 5.5 -- 8.5 \kms\ (blueshift) \\
			Downward Doppler speed in the jet spire's northern edge  & 1.2 -- 2.1 \kms\ (redshift) \\
			Spire's head-on spin direction   &  clockwise \\
			\noalign{\smallskip}\tableline\tableline \noalign{\smallskip}
		\end{tabular}
\end{table*}		
	\end{center}


%

%
\section{Results}
\subsection{Overview of the Small-scale Jet}\label{over}
\begin{figure*}
	\centering
	\includegraphics[width=\linewidth]{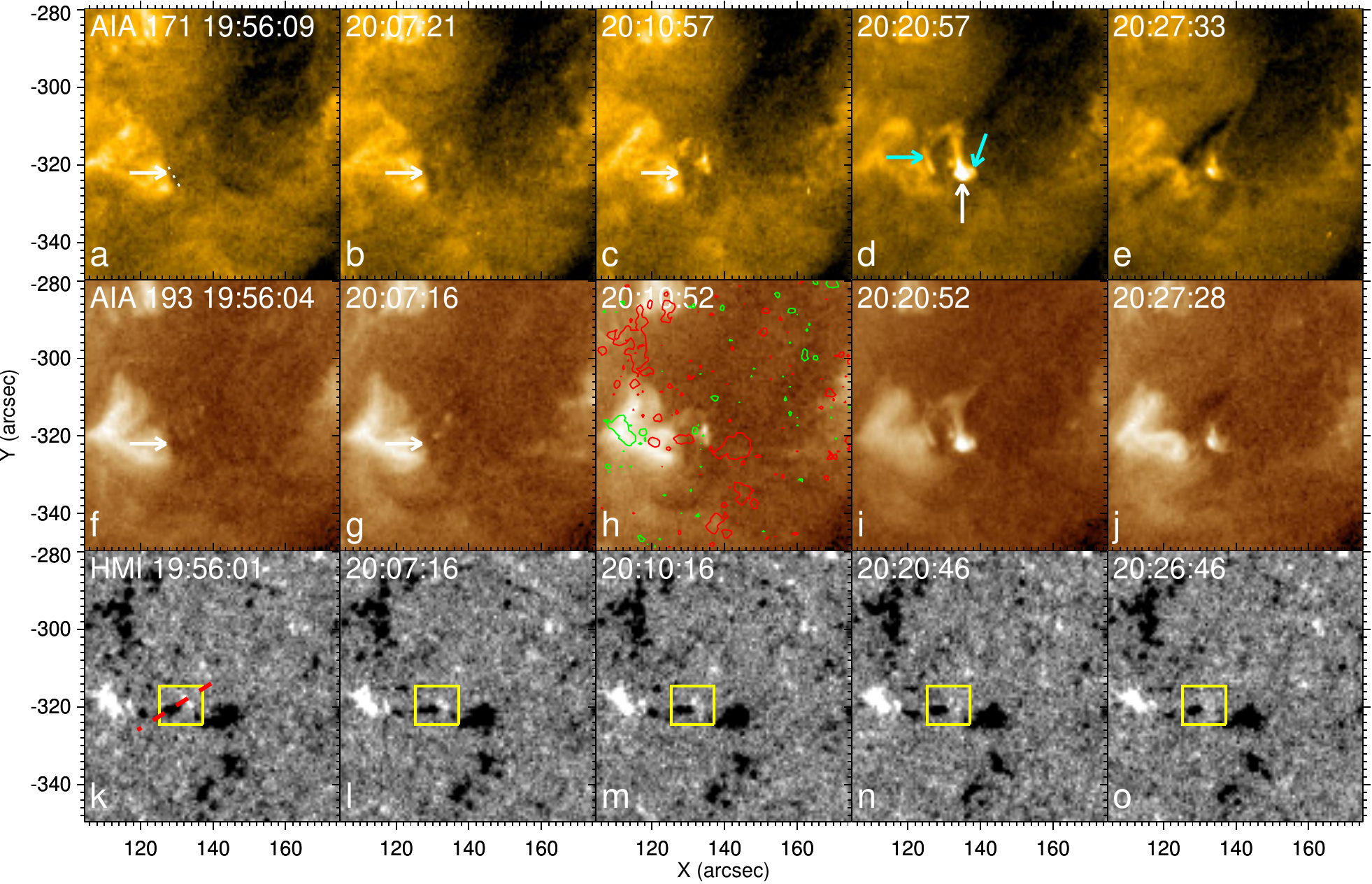}
	\caption{Formation and jet-generating blowout eruption of the minifilament in AIA images, and the canceling magnetic flux at the underlying magnetic neutral line in HMI magnetograms. Panels (a-e) and (f-j) are, respectively, 171  \AA\ and 193 \AA\ images of the  minifilament and jet and have the  FOV of  the white box in Figure \ref{fig1}. Panels (k-o) are HMI line-of-sight (LOS) photospheric magnetograms of the same FOV. The white horizontal arrows in (a,b,c,f, and g) point to the minifilament, which resides above the magnetic neutral line before the eruption. The cyan and white arrows in (d) point, respectively, to the remote brightening and the base-edge JBP brightening that appear along with the minifilament eruption. In panels (k-o), the yellow box is the area in which the  flux is measured for the time-flux plot in Figure \ref{fig3}. The red dashed-line, in panel (k), shows the diagonal cut for the time-distance image in Figure \ref{fig3}a.
		In panel (h), HMI contours, of levels $\pm$20 G, at 20:10:16 UT are overlaid, where green and red contours outline positive and negative magnetic flux, respectively. The animation runs from 19:30 UT to 20:59 UT. The AIA 193 \AA\ and HMI frames of the movie have the same annotations as in the figure. 
	}  \label{fig2}
\end{figure*}

Figure \ref{fig1} shows the coronal jet in an AIA 211 \AA\ image and  an \iris\ SJI in the 1400 \AA\ filter. We chose to show an AIA 211 \AA\ image because the coronal hole is best seen in this channel. The jet occurs in a coronal hole region in which negative-polarity flux is dominant. It is rooted at the edge of a lane of negative-polarity magnetic flux, between that majority-polarity (negative) flux  and a smaller patch of minority-polarity (positive) flux (Figures \ref{fig2}k,h). As mentioned above, this jet is fully covered by the \iris\ slit raster. The two vertical dashed lines in Figure \ref{fig1} mark the region that is scanned by the \iris\ slit for about five hours. The jet peaks around 20:20 UT.  We analyze the jet region from 19:30 UT to 21:00 UT to cover the pre-jet and jet evolution well before and after the jet erupts. The white box is the FOV that we analyze in detail and show in most other images (Figures \ref{fig2}, \ref{fig4}, \ref{fig5}, and \ref{fig6}). 

 From each raster scan, we made a spectroheliogram at the core of the \MgII\ k line, at 2796.38 \AA.   \MgII\ spectroheliograms and spectra are  shown in Figures \ref{fig4} and \ref{fig5}. The \iris\ spectroheliograms clearly show the minifilament in the jet-base region, before and during the eruption.  To show the evolution of the minifilament in the \MgII\ spectra, we made  time-wavelength maps at a different location along the slit at each of four close-together slit locations (Figures \ref{fig4}f-i). We also inverted the  \iris\ \MgII\ k line spectra with the  IRIS$^{2}$ code \citep{dalda19} to obtain  thermal and dynamical properties of the jet, namely,  temperature, line-of-sight velocity, and electron density (Figures \ref{fig6} and \ref{fig1ap}). Table \ref{tab:list} lists all the measured parameters of the jet and  minifilament.

%

\subsection{Minifilament and Magnetic Field}\label{aiafila}

Figure \ref{fig2}(a--e) and Figure \ref{fig2}(f--j)   show the jet in AIA 171 and 193 \AA\ images, respectively. A corresponding video (Movie1.mp4) shows the evolution of the jet for 1.5 hours. The minifilament appears as a dark structure. It first shows-up at 19:56 UT in the AIA images and gradually grows longer and wider (white arrows in Figures \ref{fig2}a,f). As described later in Section \ref{spectra}, the minifilament shows-up as a dark structure at the same time in \iris\ \MgII\ spectroheliograms as well. By $\sim$20:10 UT, it is obviously thicker. The minifilament starts to rise slowly at about 20:10 UT and later, at $\sim$ 20:18 UT, an obvious brightening (JBP in Figure \ref{fig2}d) starts just west of  the minifilament at the neutral line, before the start of the jet spire at $\sim$ 20:20 UT. The JBP is also visible in  AIA 94 \AA\ images. Flickering brightening appears and disappears at the same location before the start of the relatively brighter JBP.
The erupting minifilament-carrying flux rope slowly  becomes wider and moves towards the northwest until 20:20 UT and then more rapidly undergoes its blowout eruption. Figures \ref{fig2}d,e show this is a blowout eruption because the width of the bright and dark spire in Figure \ref{fig2}(e) is comparable to the width of the jet base \citep{moore10,moore13,sterling22}.
Note that the AIA 171 \AA\ images at and near the time of Figure  \ref{fig2}e also show a much fainter possible jet spire that extends southeast from near the base of the main jet spire.  The faint possible jet spire seen in AIA 171 \AA\ images does not show at all in images from AIA’s relatively hotter 193, 211, and 94 \AA\ channels.  Only the production of the much more robust main jet spire extending to the northwest is the focus of this paper.  Except for pointing it out here, we ignore the faint possible jet spire extending to the southeast.
The jet spire extends outward with an average speed of 30 $\pm$ 4 \kms.
As the JBP approaches maximum brightness (Figures \ref{fig2}d,i),  some remote brightenings appear to the east and west of the minifilament (see cyan arrows in Figure \ref{fig2}d). The western lobe brightening (downward cyan arrow in Figure \ref{fig2}d) starts at about 20:11 UT  and eastern remote brightening (horizontal cyan arrow in Figure \ref{fig2}d) starts at about 20:20 UT. 
The total duration of the jet spire is 10 minutes (the time from when the spire starts until it fades out in 171 \AA\ images).



\begin{figure}
	\centering
	\includegraphics[width=.83\linewidth]{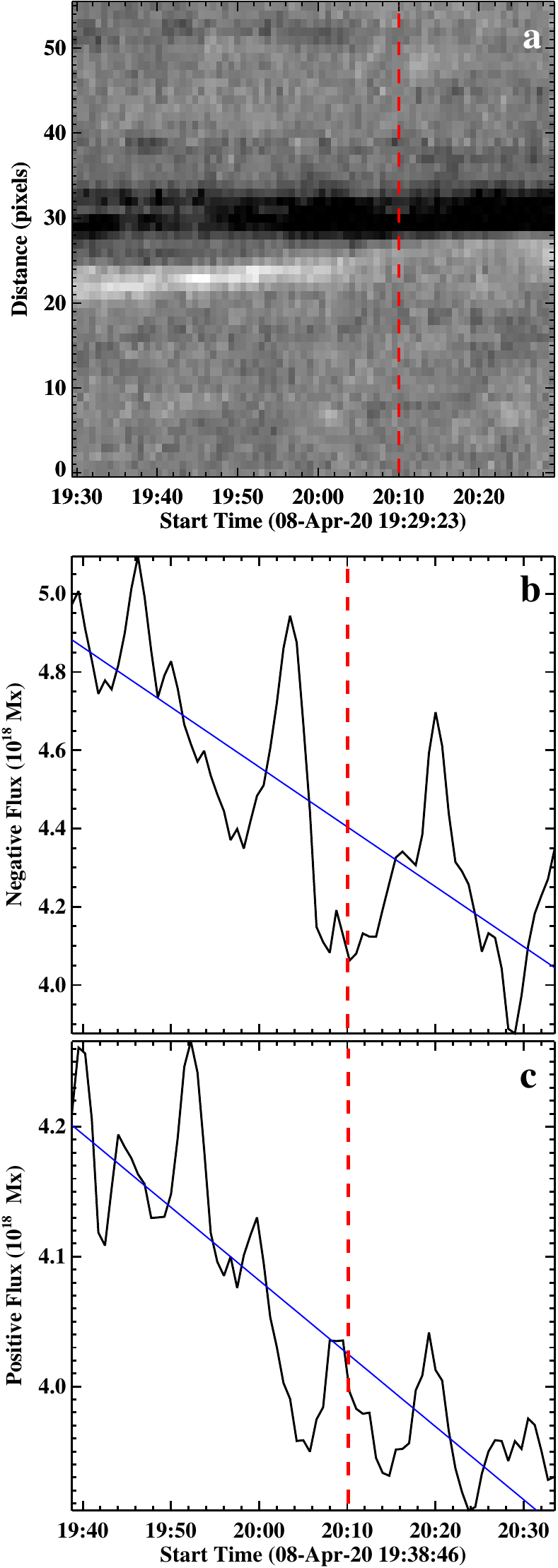}
	\caption{Canceling magnetic flux at the magnetic neutral line under the minifilament. Panel (a) is the HMI LOS time-distance image along the red dashed line of Figure \ref{fig2}k. Panels (b) and (c), respectively, display the decline of the jet-base absolute negative and positive magnetic flux with time before and through the eruption. The plotted negative and positive magnetic flux are that measured in the yellow box in Figure \ref{fig2}. The red-dashed line marks the slow-rise start time of the minifilament. The blue line is the  least-squares fit to the plotted negative or positive flux.
	}  \label{fig3}
\end{figure}
\begin{figure*}
	\centering
	\includegraphics[width=\linewidth]{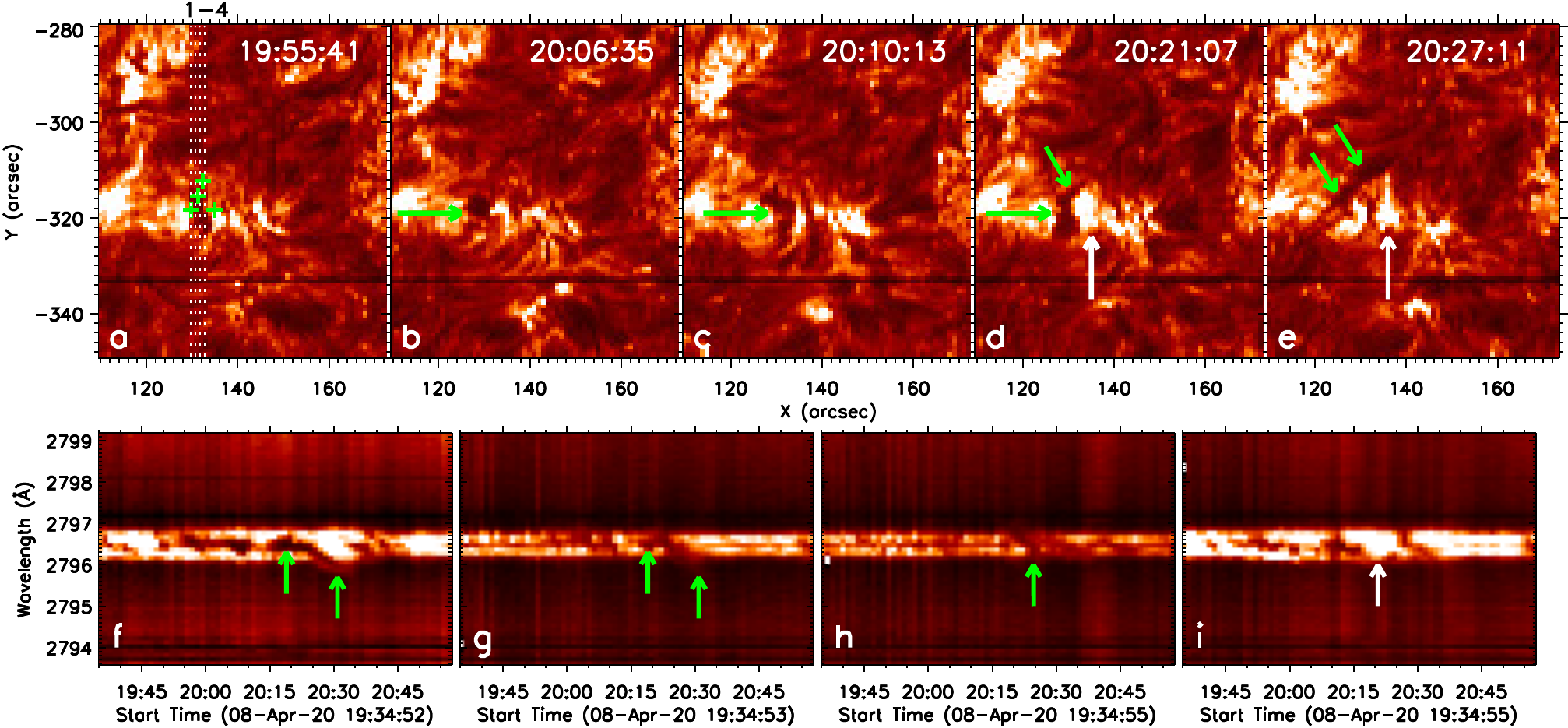}
	\caption{Evolution of the minifilament and development of the jet spire in \MgII\ line. Panels (a-e) show the spectroheliograms of \MgII\  k line at 2796.38 \AA. Each of these rasters cover the FOV bounded by the two vertical dashed lines in Figure \ref{fig1}. In (b--e) the horizontal green arrows point to the minifilament and the slanted green arrows point to the jet spire. Panels (f-i) show the temporal evolution of the spectra at four different locations marked by a `+' sign in panel (a). These `+' signs are taken at four different pixels, along the \iris\ slit, crossing the jet at four places. Panels (f), (g), (h), and (i) are along the first, second, third, and fourth `+' signs, respectively (from left to right). 
		The green arrows point to the evolving minifilament. The white arrows point to the JBP jet-base brightening. In panel (a), the dotted-white lines show the four slit positions (labeled 1--4 on top of the image) along which the full spectra of the  \MgII\  k line are shown in Figure \ref{fig5}. The panels (a--e) are taken from the animation. The animation runs from 19:30 UT to 20:59 UT and it is unannotated. 
	} \label{fig4}
\end{figure*}
\begin{figure*}
	\centering
	\includegraphics[width=\linewidth]{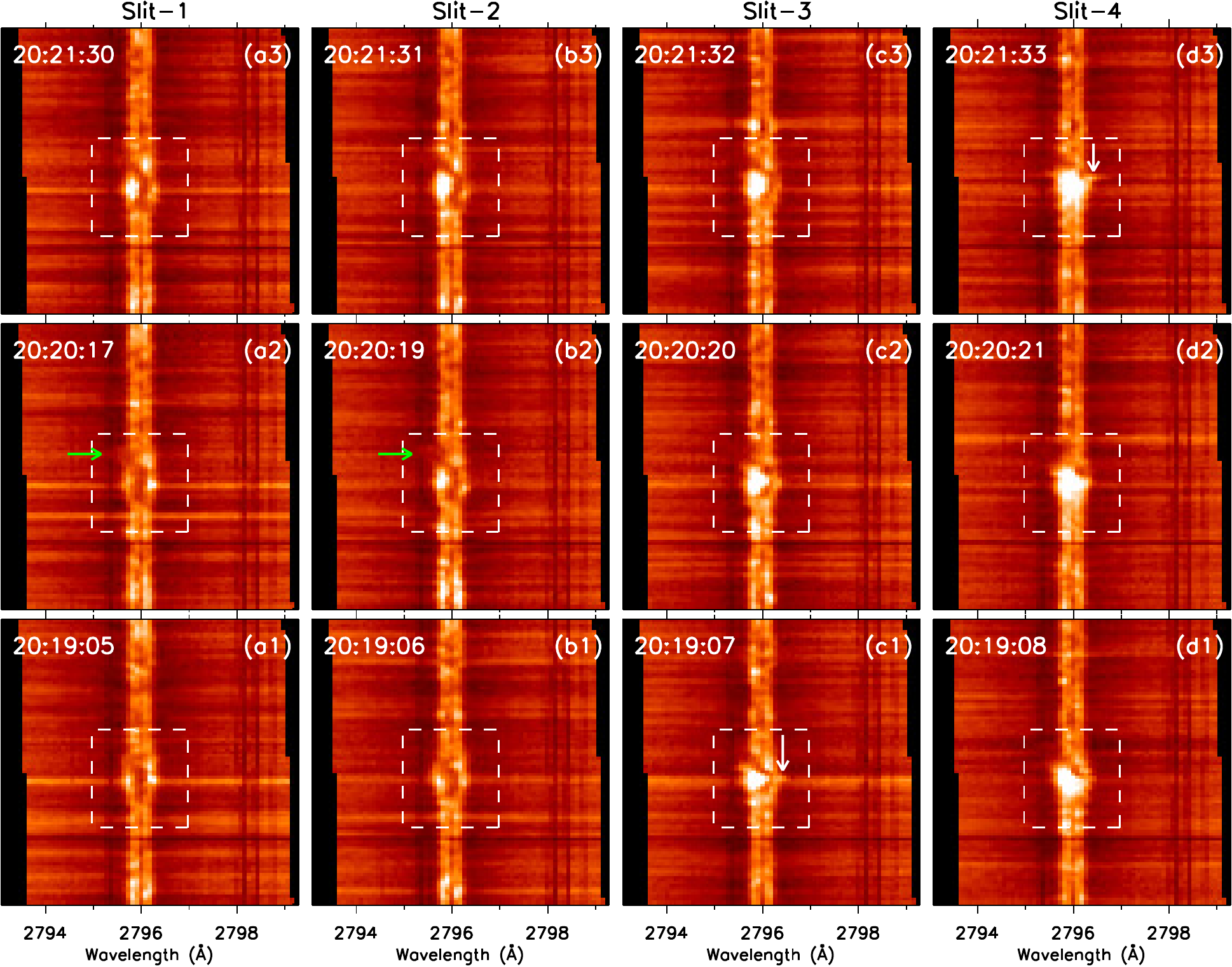}
	\caption{Progression of the minifilament eruption in \iris\ spectra (from bottom to top). Panels (a1-a3), (b1-b3), (c1,c3), and (d1-d3) show the spectra in \MgII\ k line along the four slit positions, of Figure \ref{fig4}a, at three times. The slits 1, 2, and 3 cross the minifilament and the jet spire, and slit 4 crosses the JBP. The white dashed boxes are centered on the minifilament and JBP. In d1-d3, the center of the JBP brightens under the erupting minifilament. The green arrows (in a2 and b2) point to the shift of the minifilament towards shorter wavelength, whereas the white arrows (c1 and d3) point to the extension of the JBP towards longer wavelength. } \label{fig5}
\end{figure*}

Figures \ref{fig2}(k--o) show the photospheric magnetic flux in  the jet-base region. The jet occurs at an edge of a lane of flux of the negative-polarity network. The minifilament lies above the magnetic neutral line between the lane of majority-polarity (negative) flux  and a  small patch of minority-polarity (positive) flux  (Figure \ref{fig2}h) that merges with it. 
We made a time-distance image (Figure \ref{fig3}a) along the red dashed line of Figure \ref{fig2}k, to see the magnetic field evolution at the neutral line and notice that positive flux converges and cancels with the negative network flux at the neutral line (also see yellow box in Figure \ref{fig2}). Eventually the minifilament forms/appears at the canceling neutral line and the JBP also forms at the cancelation  location, i.e. on the neutral line on which the filament sat before it erupted (Figure \ref{fig2}). Because the minifilament forms/appears at the canceling neutral line, we interpret that this minifilament forms via the flux cancelation process suggested by  \cite{balle89} for typical solar filaments.


%
%
%

In Figure \ref{fig3}, we show the gradual decline with time of  the total negative-polarity flux and total  positive-polarity flux. Both plotted magnetic fluxes are measured inside the yellow box in Figure \ref{fig2} and they indicate that magnetic flux cancels at the neutral line. The bumps in the flux plots are due to the noise level in the magnetograms as we are looking near the detection limit of the HMI magnetograms. Nonetheless, the overall trend in both polarities supports magnetic flux cancelation.

The plot supports our interpretation that flux cancelation takes place at the neutral line before and during the time of the jet, and that cancelation presumably triggers the minifilament eruption. The positive flux drops with a rate of 0.33 $\times$ 10$^{18}$ Mx hr$^{-1}$ whereas negative flux drops with a rate of 0.9 $\times$ 10$^{18}$ Mx hr$^{-1}$.  Movie1 suggests that the decrease in negative flux is roughly three times the decrease in positive flux because some negative flux leaks out the east side of the box and some leaks out the southwest coroner of the yellow box of Figure \ref{fig3}a. Note that this is not the case of unipolar magnetic flux disappearance due to dispersion/fragmentation as found in some magnetic features by \cite{anusha2017}. In our case there is a clear  flux convergence and disappearance of both polarities together as is evident from the magnetic flux plots (Figure \ref{fig3}b,c) as well as from Movie1. We interpret that this is flux cancelation via reconnection driven by converging photospheric flows as suggested by, among others, \cite{balle89,moore92,tiwari14,kaithakkal19,syntelis21,priest21,hassanin22}. 

It is important to note that even though we were able to isolate each magnetic polarity (negative polarity to  some extent) in this case, often only one polarity flux can be properly isolated. In those cases it is accepted that magnetic flux measurement can be made for only one polarity and a reduction in the flux in that polarity alone is evidence of ongoing flux cancelation; see \cite{green11} for further details.


%

%
%
%
\subsection{Mg II Spectroscopic Observation of  Minifilament }\label{spectra}

During and before the jet eruption, the \iris\ slit raster covered the minifilament and jet spire thus enabling us to perform our spectroscopic analysis of the erupting minifilament and  jet. \iris\ observed, in the \MgII\ k resonance lines, the strong emission and absorption at the location of JBP and minifilament, respectively.

Figures \ref{fig4}(a--e) and Movie2.mp4 show the spectroheliograms of the \MgII\ k line that are made at 2796.38 \AA. The horizontal green arrows point to the evolving dark minifilament structure in the jet base region whereas slanted green arrows point to the jet spire. In accord with the AIA images, the erupting minifilament starts becoming the dark spire at 20:20 UT and after that it continues to escape into the spire as the spire shoots out in the north-west direction (see green arrows in Figure \ref{fig4}e). Finally, it is fully ejected into the spire at 20:30 UT.  The white arrows in Figure \ref{fig4} point to the JBP that grows underneath the erupting filament. The \MgII\ minifilament appears similar to that reported by \cite{hermans86,wang00} in \Halpha.

In Figures \ref{fig4}(f--i), we show the temporal evolution of the spectra at  the `+' signs of Figure \ref{fig4}a. These `+' signs are  at four different locations along  the \iris\ slit, three on the minifilament and jet spire, and one on the JBP. The first map of Figure \ref{fig4}f is from the first `+' sign of Figure \ref{fig4}a (from left). Similarly, Figures \ref{fig4}g,h,i are from the second, third and fourth `+' signs, respectively, starting from the left side. The minifilament activity can be seen between 20:15 and 20:30 UT (see green arrows), which is consistent with the activity seen in the spectroheliogram video (Movie2). The time-wavelength map of the JBP location is plotted in Figure \ref{fig4}i. One can see strong emission that appears at the location of the JPB as the minifilament erupts.

We also track the evolution of the minifilament in the \MgII\ k line spectra, along the four different slit positions marked in Figure \ref{fig4}a. The slit spectra of the minifilament and the JBP are displayed in Figure \ref{fig5}, at three times. The  spectra along the slit position `1' show the Doppler shifts of the minifilament as the JBP turns on (see inside the white boxes of Figure \ref{fig5}a1--a3). One can see that the minifilament gets slightly shifted towards the shorter wavelength side (e.g. in Figure \ref{fig5}a2 as compared to Figure \ref{fig5}a1), which means that the minifilament gets blue-shifted during the eruption. Similar blueshifted upflows can also be seen in the minifilament in the spectra from slit position `2' (green arrow in Figure \ref{fig5}b2). 
Similar spectra for a minifilament in a coronal bright point has recently been reported by \cite{madjarska22}. However they did not study the dynamics of the minifilament due to the lack of repeated IRIS raster scans.
The spectra of slit position `3' and `4' show the evolution of the JBP. When the JBP forms under the erupting minifilament, the spectra shift towards the red. 
This redshift shows there are downflows at the location of the JPB (see white arrows in Figure \ref{fig5}c1 and Figure \ref{fig5}d3). As discussed in Section \ref{inversion}, this downflow could be caused by magnetic reconnection as a result of which a part of the material moves up and some moves down.


\subsection{Inversion of \MgII\ spectra}\label{inversion}
\begin{figure*}
	\centering
	\includegraphics[width=\linewidth]{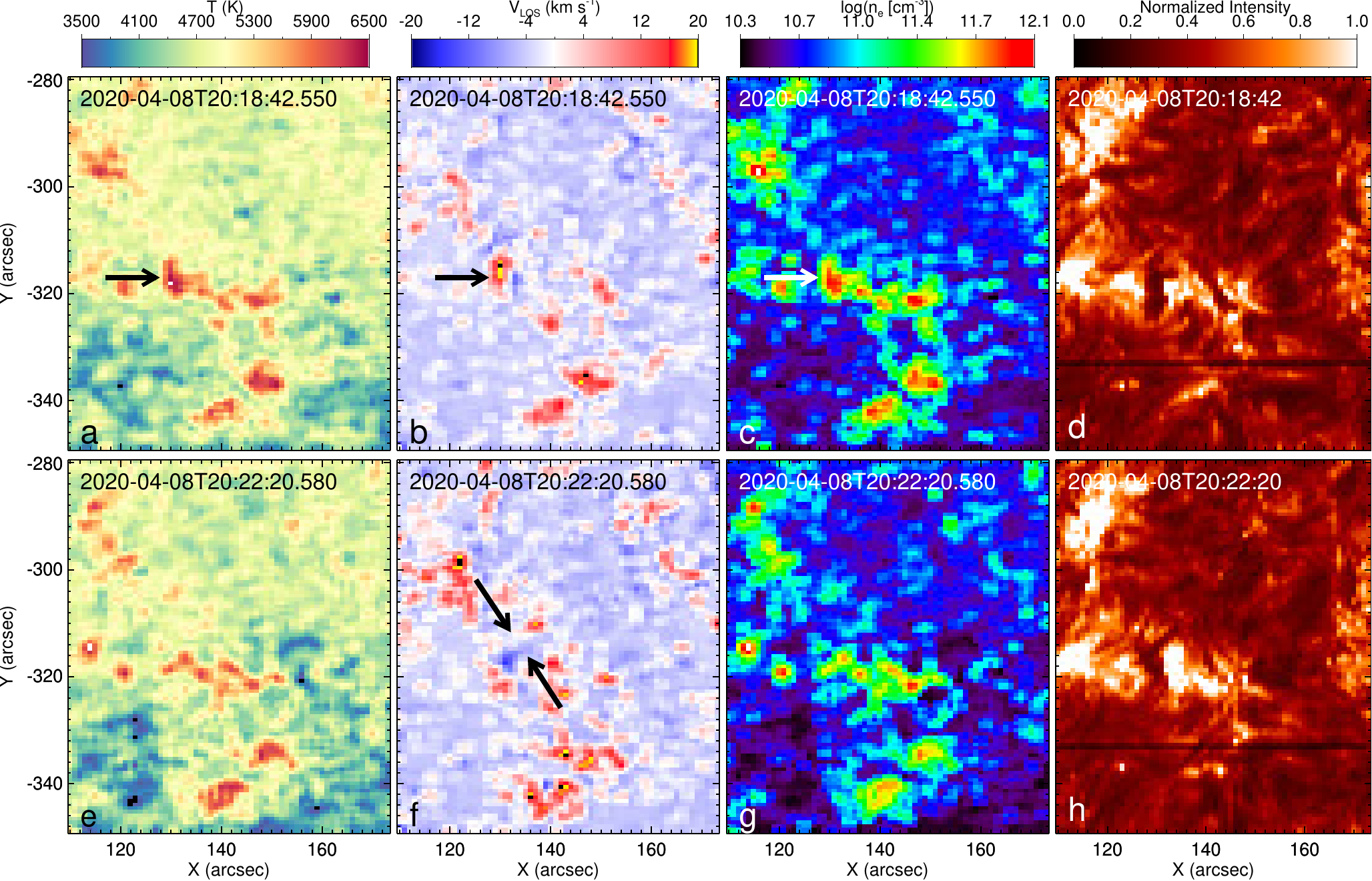}
	\caption{Thermal and dynamical properties of the minifilament eruption. Panels (a, e), (b, f), and (c, g) show the temperature, line of sight velocity (V$_{LOS}$) and electron density (log(n$_{e}$)) maps, respectively, evaluated at log($\tau$) =-4.2. Panels (d, h)  show the spectroheliograms of \MgII\  k line at 2796.38 \AA. The black and white arrows in (a--c) point to the JBP. In panel (f) the upper black arrow points to the weak redshifts and the lower arrow points to the blueshifts  along the jet spire.
	}  \label{fig6}
\end{figure*}

To map  the thermal and dynamical properties of the minifilament eruption and jet, we applied the 
IRIS$^{2}$ inversion code to the \iris\ \MgII\ spectra. The details of the IRIS$^{2}$ inversion code are available in the papers by \cite{dalda19} and \cite{rodriguez19}, and at the LMSAL webpage\footnote{https://iris.lmsal.com/iris2/}. Figure \ref{fig6}  shows, during the jet onset, the inverted temperature, line of sight velocity (V$_{LOS}$)  and electron density log(n$_{e}$)  maps of the same FOV that is shown in Figure \ref{fig4}(a--e).  Those maps are for log($\tau$) =-4.2.  We also looked at maps for  log($\tau$) =-4.0, -4.4, -4.6, -4.8, and -5.0 (the maps for log($\tau$)=-5.0 are shown in Appendix Section  \ref{appen}). The maps  show the strongest Doppler signal (JBP redshift and minifilament blueshift) at log($\tau$) =-4.2.  During the slow-rise phase of the minifilament eruption (from 20:10 to 20:19 UT) some parts show varying weak blueshifts of  no more than 6 \kms. 


The arrows in Figures \ref{fig6}(a--c) point to the location of  the JBP under the  rising minifilament (Figure \ref{fig6}d). At that location, we observe enhancement in temperature and density maps. We interpret that the increase in density and temperature is due to the reconnection brightening (i.e. JBP), which forms underneath the minifilament. The inversion maps show  downward (red-shift, Figure \ref{fig6}b) Doppler speed of 10 \kms, and that the  JBP’s \MgII\ bright plasma has electron density and  temperature of   10$^{12}$ cm$^{-3}$ and 6000 K, respectively. We interpret that  the redshifts result from downward flow from the reconnection above. These values correspond to the cool material in the JBP, because the \MgII\ line is only formed at chromospheric temperatures \citep{rodriguez19}.  Because the JBP emits in AIA 211 and  94 \AA, it is clear that it also contains multi-million degree coronal-temperature plasma  \citep{lem12}.  The \MgII \ channel however does not detect emission from those plasmas.

As mentioned above and shown in Figure \ref{fig6}h, the minifilament-carrying flux rope erupts  towards the solar north-west. During that stage, we observe upward (blueshift, Figure \ref{fig6}f) Doppler speeds in the range of 5.5 to 8.5 \kms\ with an uncertainty of $\pm$ 1.5 \kms. The Dopplergram also shows the signature of weak red-shifts (1.2--2.1 $\pm$ 1.0  \kms) just next to the blueshifts (see black arrows in Figure \ref{fig6}f). Registration of the Dopplergram in Figure \ref{fig6}f with the \MgII\ spectroheliogram in Figure \ref{fig6}h shows that the two opposite Dopplershits are on the opposite edges of the jet spire seen in Figure \ref{fig6}h.	
The opposite line-of-sight velocities along opposite edges of the jet spire crossed by the \iris\ slit is evidence that the jet spire is spinning. Viewed from its top (from the north-west toward the south-east), the jet spire is spinning clockwise about its axis, consistent with clockwise untwisting of the magnetic field in the jet spire. The evidence for the clockwise direction of the spire's twist is that when viewed from the north-west, the spire’s blue-shifted part seen in the \iris\ \MgII\ spectra is on the left and the weak red-shifted part is on the right. 

We note that the redshift is not as strong in this case as the blueshift. Nonetheless, the redshift is slightly above the estimated uncertainties from the IRIS$^{2}$ inversion. Similar, but stronger, blueshifts and redshifts across the jet spire are also observed by \iris\ previously in active region jets \citep{cheung15} and in large penumbral jets \citep{tiwari18} and have been inferred to show twisting/untwisting motions. 



%
\begin{figure*}
	\centering
	\includegraphics[width=\linewidth]{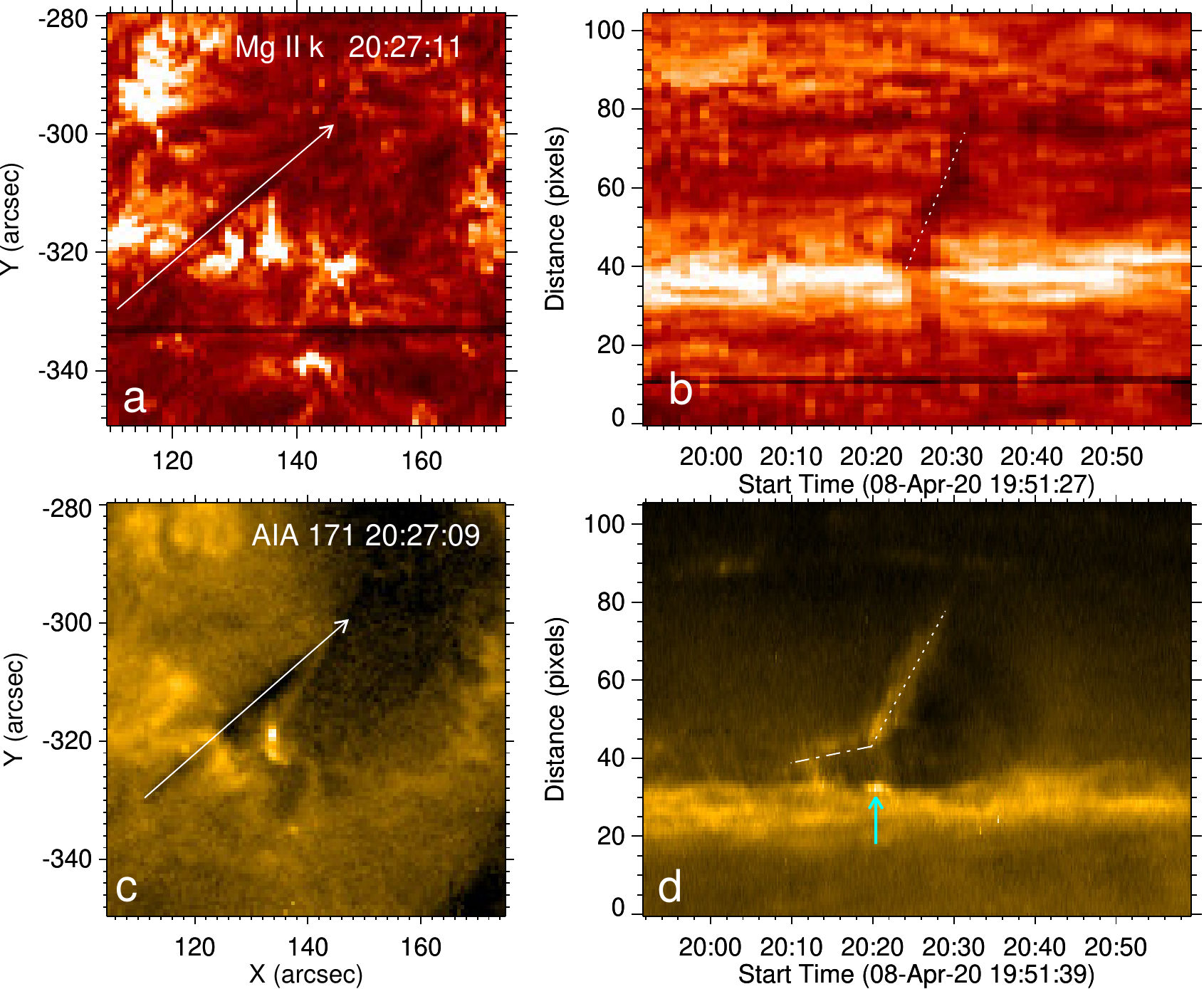}
	\caption{Jet speed. Panels (a) and (c) show a spectroheliogram of the \MgII\  k line and an AIA 171 \AA\ image of the jet. The diagonal arrows in (a) and (c) show the cut for the time-distance maps plotted in (b) and (d). Panels (b) and (d) show the time-distance maps from the cuts in  Figures \ref{fig7}a and \ref{fig7}c. Map (b) shows the progression of the tip of the dark jet spire in panel (a); map (d) shows the progression of the bright tip of the dark jet spire in panel (c). The dotted lines in (b) and (d) are the paths used to calculate the speed of plasma outflows along the jet spire. The dotted-dashed line in (d) marks the upper edge of the dark minifilament during the minifilament's slow rise. The cyan arrow in 
			(d) points to the remote brightening that appears to the east of the erupting minifilament.
	} \label{fig7} 
\end{figure*} 

\subsection{Jet Speed}
We measured the plane-of-sky speed of the jet in the \MgII\ spectroheliograms and AIA 171 \AA\ images as it erupts outward. 
In Figures \ref{fig7}b and \ref{fig7}d, we show the time-distance maps from the diagonal cuts in Figures \ref{fig7}a and \ref{fig7}c, respectively.  These cuts are along the spire of the jet. The  \MgII\ time-distance map (Figure \ref{fig7}b), shows the track of the tip of the dark spire. The 171 \AA\ time-distance map shows the trace of the bright tip of the dark spire seen in the 171 \AA\ images. To obtain the speed, we measured the slope of the dotted straight lines drawn along the spire-tip track in each of the two time-distance maps.


The slope of the dotted line in Figure \ref{fig7}b gives  32 $\pm$ 4 \kms\ for the speed of the spire's tip  in the \MgII\ spectroheliograms. The slope of the dotted line in Figure \ref{fig7}d gives 30 $\pm$ 4 \kms\ for the speed of the spire's tip in the AIA 171 \AA\ images. We also estimate the speed of the minifilament during its slow-rise phase, from the slope of the dotted-dashed line; that gives to be 3.0 $\pm$ 0.8 \kms, which is in the range of those speeds reported by \cite{panesar20a} for on-disk quiet Sun jets  (we notice that the minifilament slow-rise phase was not discernible in the \MgII\ spectroheliograms, possibly due to its coarser cadence). We repeated the measurements three times, at three different nearby locations along the jet outflows (Figures \ref{fig7}b and \ref{fig7}d) and then calculated the average and standard deviation of the three measurements. 
That the \iris\ slit size is 0.66\arcsec\ whereas the slit-step size is 1\arcsec\ was taken into account in the speed measurements from the \MgII\ spectroheliograms.  The observed plane-of-sky speeds are $\sim$ 5  times faster than the Doppler blueshift speeds as shown in Figures \ref{fig5} and \ref{fig6}. 


Our speed measurements show that the plane-of-sky speed of the jet spire's tip is  the same, within error bars, in both sets of images (\MgII\ and 171 \AA). This indicates that the \MgII\ spectroheliograms and the 171 \AA\ images show the same cool dark plasma component of the spire.


\begin{figure}
	\centering
	\includegraphics[width=\linewidth]{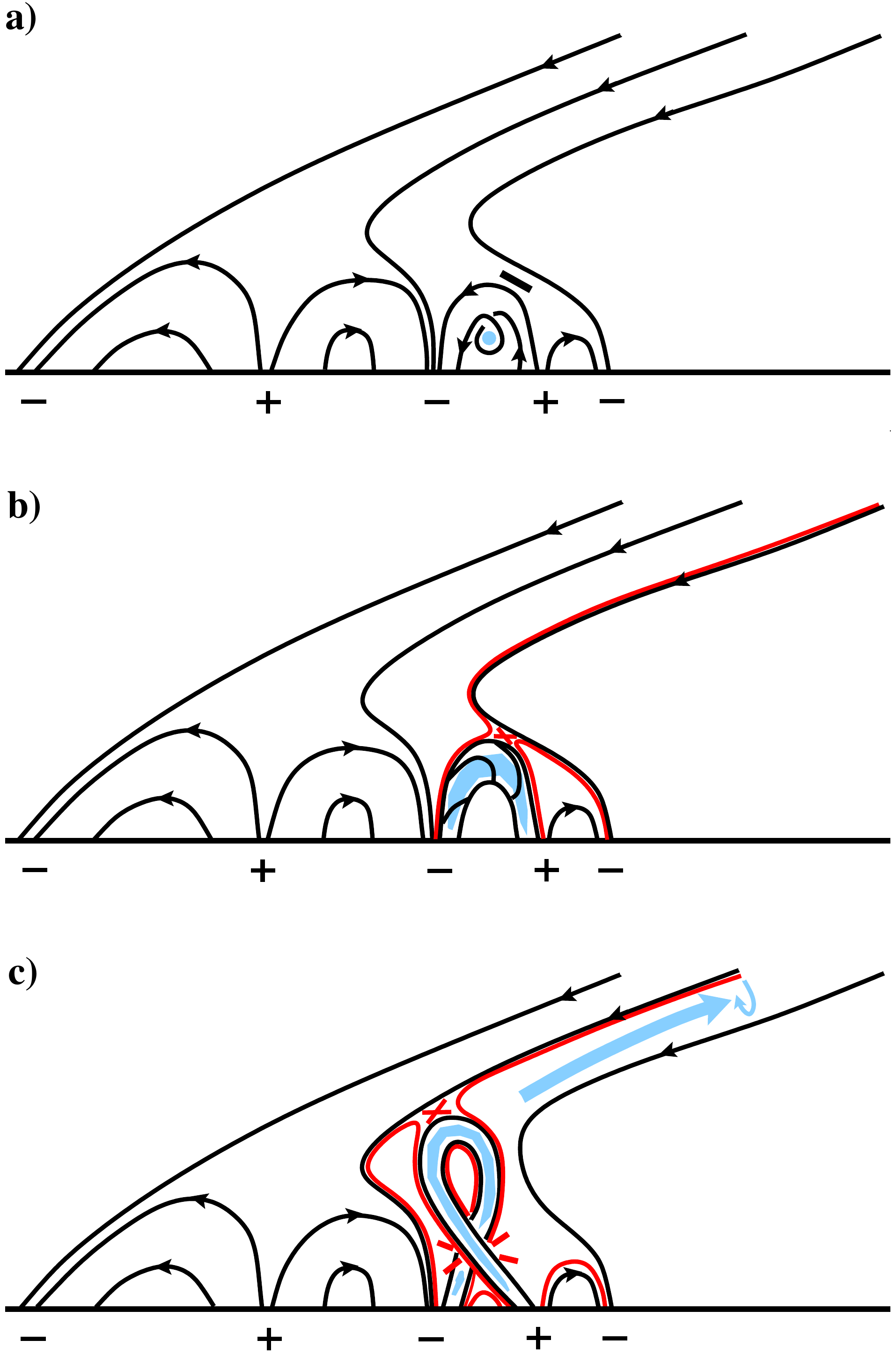}
	\caption{Schematic of the possible arrangement, action, and reconnection of the magnetic field in and	around the  minifilament before and during the minifilament-carrying field’s jet-generating blowout eruption. The view direction is from the southwest, more or less along the direction of the flux-cancelation PIL underlying the minifilament seen in Figure \ref{fig2}(a) and more or less orthogonal to the jet spire seen in Figure \ref{fig2}(e). The horizontal black line is the solar surface. Positive and negative magnetic flux polarities are labeled with ``+'' and ``--'' signs, respectively. Each red X marks a site of ongoing  magnetic field	reconnection. The minifilaments’s cool plasma and reconnected magnetic field lines are sky blue and red, respectively. The thick sky blue arrow in panel (c) is the jet spire and the elliptical arrow shows the clockwise rotation (the head of the elliptical arrow is closer to the viewer than the tail). Panel (a) shows the set-up before the eruption. The black slanted bar marks the location of current-sheet
			formation at the magnetic null point between the flux-rope-enveloping lobe and ambient far-reaching
			field. Panel (b) is early in the  slow-rise of the erupting minifilament-carrying flux rope and enveloping lobe, when
			reconnection of the outside of the lobe has started but the erupting flux rope as not yet kinked.  In panel (c), the erupting minifilament flux rope has kinked (writhed)
			to give some of its right-handed magnetic twist to a right-handed writhe. 
}  \label{skt}
\end{figure} 
\section{Discussion}

We have investigated the dynamics and magnetic flux evolution of a central-disk jet eruption in a coronal hole region using \iris\ \MgII\ k spectra, EUV images from \sdo/AIA, together with the line of sight \sdo/HMI magnetograms. This is the first detailed \iris\ \MgII\  spectroscopic analysis of the genesis and evolution of an on-disk pre-jet minifilament. The \iris\ \MgII\ k spectra and spectroheliograms show (i) a minifilament that contains plasma of chromospheric temperatures similar to the \Halpha\ filaments; (ii) the minifilament starts to form in the jet-base region 10--20 minutes  before erupting to make the jet; and (iii) blueshifted upflows in the minifilament's plasma during the eruption and concurrent  redshifted downflow at the JBP that forms underneath the erupting minifilament. AIA images and HMI  magnetograms show that the jet occurs at the edge of a clump of negative-polarity  network magnetic  flux. The jet is seated on the neutral line between the negative network flux clump and a merging small patch of positive flux. The  flux cancelation at the neutral line prepares and triggers the minifilament eruption and the blowout eruption of the minifilament flux rope drives the production of the  jet. During the eruption onset, the  JBP forms on the neutral line on which the minifilament sat prior to its eruption. These results are the same as for larger coronal-jet-generating minifilament eruptions.  Furthermore, IRIS${^2}$ inversion of the \MgII\ spectra shows that during the eruption onset the  JPB  has electron density, temperature, and downward (redshift) Doppler speed of  10$^{12}$ cm{$^{-3}$}, 6000 K, and 10 \kms, respectively. 


Figure \ref{skt} is a schematic of the blowout jet eruption, a schematic that is consistent with our \iris, AIA, and HMI observations. It shows that the minifilament (shown in sky blue color) sits above the sheared-field neutral line. The cool minifilament plasma is suspended in a twisted flux rope in the sheared-field core of the jet-base magnetic anemone’s lobe that envelops the flux-cancelation PIL. The core field and flux
rope have right-handed shear and twist (Figure \ref{skt}a). Persistent flux cancelation at the neutral line eventually destabilizes the sheared field/flux rope that holds the cool-plasma minifilament, and that field loop erupts outward. The outer envelope of the erupting flux rope reconnects with the ambient far-reaching magnetic field at the magnetic null point  (Figure \ref{skt}b) via breakout reconnection \citep{antiochos98,wyper17,wyper18}. This reconnection adds new closed field loops on the west side of the positive flux clump. We observe brightening, shown in Figure \ref{fig2}d, in the western lobe of the jet-base anemone. As the  minifilament-carrying flux rope continues to erupt, we speculate that it kinks (writhes) so that the top of the kink loop rotates clockwise.  Clockwise rotation of the top of the kink loop results from the right-handed twist in the flux rope going into right-handed writhe twist, which results in the positive-foot leg rotating in front of the negative leg (Figure \ref{skt}c). 
 Such kinking and/or writhing in large-scale erupting flux ropes has been previously reported in observations and modelling \citep[e.g.] []{torok05}.
 Magnetic reconnection occurs between the legs of the writhing field (at the lower X in Figure \ref{skt}c), which results in a brightening (the JBP that we observe in the AIA images and \MgII\ spectra; Figures \ref{fig2}d and \ref{fig4}d) at the neutral line.  Due to the writhing  (kinking), the direction of the magnetic field in the top of erupting flux-rope becomes opposite to that of overlying (presumed open) far-reaching ambient magnetic field, which enables the top of the erupting kink to reconnect with the encountered far-reaching field at the upper X in Figure \ref{skt}c. This reconnection (1) makes a newly-reconnected loop that has one foot east of the erupting minifilament and (2) transfers right-handed twist to the newly-opened (red) field lines of the spire. The cool-plasma and hot reconnection-heated plasma escape along the newly-reconnected untwisting far-reaching field and appears as the jet spire. We also observe some faint brightening in AIA images at the foot of the western newly-reconnected small loop (downward cyan arrow in Figure \ref{fig2}d). Thus, our results are consistent with our earlier observations of typical coronal jets  \citep{panesar16b,panesar17,sterling17,panesar18a} and jetlets \citep{panesar18b,panesar19}. Those are also seen to be driven by minifilament eruptions and to have at the base a minifilament flux-rope core that is built and triggered to erupt by flux cancelation. The schematic is intended to depict the production of the robust jet spire that extends to the northwest from the jet base, the production of the JBP, and the production of the brightening  east and west of the JBP.  


The lifetime of the jet is 10 minutes, which is similar to that of coronal-hole jets  \citep{shimojo96,savcheva07,panesar18a,mcglasson19}. Our jet spire extends outward with an average speed of 30 $\pm$ 4 \kms, a third the speed of average coronal jets  \citep[e.g.][]{panesar16b,panesar18a} and jetlets \citep{panesar18b}. However, the speeds of some of  jets and jetlets measured by \cite{panesar16b,panesar18a,panesar18b} are in the range of 30--50 \kms. 

During the jet onset, we observe along the spire blueshift Doppler speeds (in \MgII\ k) in the range  5.5 to 8.5 \kms. Just next to the blueshift, the Dopplergrams also show some weak redshift Doppler speeds (in the range  1.2-2.1 \kms). The presence of both signs of Doppler shift along the jet spire could be from untwisting motion of the magnetic field in the jet spire, albeit the redshift in this case is much weaker than the blueshift. The presence of blueshift and redshift next to each other leads us to assume clockwise writhing of the minifilament flux rope as it erupts. The untwisting/spinning motion in coronal jets have been seen in many observations \citep[e.g.][]{pike98,kamio2010,curdt2012,moore15,panesar16a,sterling16,sterling19} and jet models \citep[e.g.][]{pariat15,wyper18,zhelyazkov18,wyper19,doyle19}. 

In the past, \iris\ spectra have captured stronger both side-by-side opposite Doppler shifts in active region jets \citep[e.g.][]{cheung15,lei19,ruan19,tiwari19,zhang21} and in large penumbral jets \citep{tiwari18}.  None of these studies focus on the presence and dynamics of cool-plasma structure that often appears in the jet base in typical coronal jets. In this regard, the present paper  is the first detailed  study of an on-disk pre-jet minifilament and its jet-generating eruption  using \iris's \MgII\ k line spectra. We note that capturing the spatio-temporal evolution of a coronal jet, including its jet-base region and jet spire, under the \iris\ slit is difficult due to the required high temporal cadence of the data for studying jet dynamics. Furthermore, due to the large size of normal-size coronal jets and the uncertainty of when and where they occur, often only a small portion of the jet is  captured. Future jet observations from multi-slit instruments such as MUSE \citep{pontieu20} that can scan a bigger FOV at a high temporal cadence and dense raster will give us opportunity to study a larger sample of jets  in EUV channels. The high-resolution observations of the  Swedish Solar Telescope, and DKIST, together with \iris, will shed more light on the formation mechanism of chromospheric structures such as pre-jet minifilaments. Again, however, the main challenge would remain to capture such events, like the one we studied here, during an observation, due to smaller FOV and limited raster cadence of ground-based observations. 
 
The appearance of our minifilament in the  \MgII\ line is similar to such small-scale filaments observed in \Halpha\  \citep{hermans86,wang00,lees03}:  the filaments reside above  magnetic neutral lines at edges of network flux clumps in quiet-Sun regions; they erupt during flux cancelation at the neutral line;  the  filaments stop reappearing and erupting when all of the minority-polarity magnetic flux patch has canceled. 
Our observations are in agreement with the above mentioned studies, which indicate that a  minifilament is a small-scale analog of  larger-scale filaments. The main differences are that in the above-mentioned studies \Halpha\ minifilaments were not necessarily accompanied by typical coronal jets, and that they did not include spectroscopic observations of minifilaments. Furthermore, in the present work we are able to invert the Mg II spectra using IRIS$^2$ technique to determine velocities, temperatures, and densities of the pre-jet minifilament and the JBP.


\section{Conclusion}
Using \iris\ \MgII\ rastered spectra and spectroheliograms, and \sdo\ data (AIA EUV images and HMI magnetograms), we investigate the formation of a minifilament in an on-disk coronal hole and the detailed dynamics of the minifilament's eruption that makes a coronal jet. The  minifilament forms on and erupts from a magnetic neutral line at which flux cancelation is occurring in the base of the jet. In this way, the eruption of the observed small-scale coronal jet appears to  work in  the same way as the blowout minifilament eruptions that make many other coronal jets. We suppose that, a la \cite{balle89}, the flux cancelation gradually builds up shear in the magnetic field that holds the minifilament of cool plasma and makes a minifilament-carrying flux rope from that sheared field, and that, a la \cite{moore92} the flux cancelation eventually triggers the flux rope’s blowout eruption.

To our knowledge, our coronal jet is the first whose dynamics is fully captured in rastered \iris\ spectra and analyzed in detail. This \iris\ jet shows all of the characteristics of typical coronal jets. In particular, we have found direct evidence of flux cancelation at the underlying neutral line forming a sheared field/flux rope holding a minifilament and triggering that core field's jet-generating eruption. Some minifilament plasma is blueshifted during the eruption, while some is redshifted in down-flow at the JBP. We expect that similar observations, including UV spectra, of jets of all sizes, including  jetlets and jet-like campfires \citep{panesar2021}, will further clarify jet eruption mechanisms, the similarity of jets of all sizes, and whether and how network jets and jetlets might drive coronal heating.



\vspace{5cm}

We thank the referee for useful comments that improved the  presentation of the manuscript. NKP acknowledges support from NASA’s HGI, SDO/AIA, and HSR  grants.  SKT gratefully acknowledges support by NASA's HGI and Hinode grants. RLM and ACS were supported by funding from NASA's HGI and HSR grants. BDP was supported by NASA contract NNG09FA40C (IRIS).
We acknowledge the use of  \iris\ and  \sdo/AIA/HMI data. AIA is an instrument onboard the Solar Dynamics Observatory, a mission for NASA’s Living With a Star program. \iris\ is a NASA small explorer mission developed and operated by LMSAL with mission operations executed at NASA Ames Research center and major contributions to downlink communications funded by ESA and the Norwegian Space Centre. This work has made use of NASA ADSABS. We thank Alberto Sainz Dalda for discussions on IRIS$^{2}$ inversion code. 


\bibliographystyle{aasjournal}

\appendix

\section{Appendix Information}\label{appen}

\begin{figure*}[h]
	\centering
	\includegraphics[width=\linewidth]{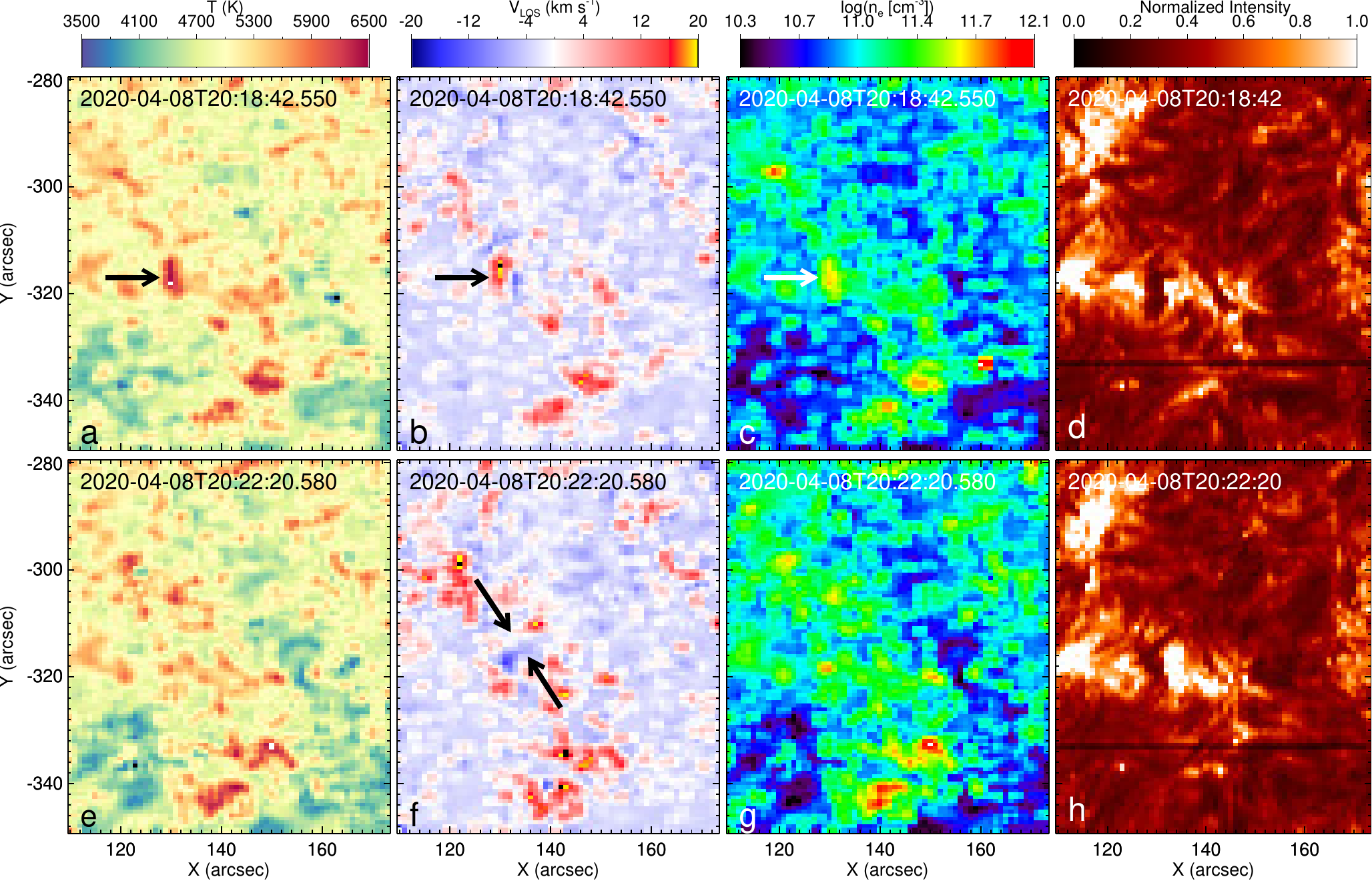}
	\caption{Thermal and dynamical properties of the minifilament eruption.  Panels (a, e), (b, f), and (c, g) show the temperature, line of sight velocity (V$_{LOS}$) and electron density (log(n$_{e}$)) maps, respectively, evaluated at log($\tau$) =-5.0. Panels (d, h)  show the spectroheliograms of \MgII\  k line at 2796.38 \AA. The black and white arrows in (a--c) point to the JBP. In panel (f) the upper black arrow points to the weak redshifts and the lower arrow points to the blueshifts  along the jet spire; these Dopplershifts confirm the Dopplershift evidence in Figure \ref{fig6}f for clockwise spin of the spire.
	}  \label{fig1ap}
\end{figure*}

Figure \ref{fig1ap}  shows the inverted temperature, line of sight velocity (V$_{LOS}$)  and electron density log(n$_{e}$)  maps of the same time and FOV as  shown in Figure \ref{fig6}, but for log($\tau$) =-5.0.  The observed Doppler speeds are of the same order as found for log($\tau$) =-4.2. The average temperature and electron density of the JBP are 6300 K and 2.8 $\times$ 10$^{12}$ cm$^{-3}$, respectively.

\end{document}